\documentclass{ws-procs9x6}

        \newcommand{\be}{\begin{equation}}
        \newcommand{\ee}{\end{equation}}
        \newcommand{\bea}{\begin{eqnarray}}
        \newcommand{\eea}{\end{eqnarray}}

\def\tr{{\rm tr}}

\newcommand{\sdiv}{{div}}
\newcommand{\beq}{\begin{equation}}
\newcommand{\eeq}{\end{equation}}
\newcommand{\boldl}{{\bf L}}

\newcommand{\calp}{{\mathcal P}}

\newcommand{\vx}{\vec{x}}

\newcommand{\calz}{{\mathcal Z}}
\newcommand{\ellthree}{\ell_{\bf 3}}

\newcommand{\ellfund}{\ell_{{\bf N}}}

\newcommand{\wideellfund}{\widetilde{\ell}_{{\bf N}}}
\newcommand{\wideellfundconj}{\widetilde{\ell}_{\overline{{\bf N}}}}

\newcommand{\boldlN}{{\bf L}_{{\bf N}}}

\newcommand{\calr}{{\mathcal R}}

\def\beq{\begin{equation}}
\def\eeq{\end{equation}}

\def\ibid#1#2#3{{\it ibid.} {\bf #1}, #2 (#3)}
\def\ijm#1#2#3{Intl. Jour. Mod. Phys. {\bf #1}, #2 (#3)}
\def\jhep#1#2#3{JHEP {\bf #1}, #2 (#3)}

\def\npb#1#2#3{Nucl. Phys. B {\bf #1}, #2 (#3)}
\def\npsb#1#2#3{Nucl. Phys. Proc. Suppl. B {\bf #1}, #2 (#3)}
\def\plb#1#2#3{Phys. Lett. B {\bf #1}, #2 (#3)}
\def\prc#1#2#3{Phys. Rev. C {\bf #1}, #2 (#3)}
\def\prd#1#2#3{Phys. Rev. D {\bf #1}, #2 (#3)}
\def\prl#1#2#3{Phys. Rev. Lett. {\bf #1}, #2 (#3)}
\def\phr#1#2#3{Phys. Rep. {\bf #1}, #2 (#3)}

\def\rmp#1#2#3{Rev. Mod. Phys. {\bf #1}, #2 (#3)}

\def\zpc#1#2#3{Z. Phys. C {\bf #1}, #2 (#3)}

\begin{document}
\title{Renormalized Polyakov Loops, Matrix Models and the Gross-Witten Point}

\author{Adrian Dumitru}

\address{Institut f\"ur Theoretische Physik, J.W.~Goethe Univ.,\\
	Postfach 11 19 32, 60054 Frankfurt, Germany  \\ 
	E-mail: dumitru@th.physik.uni-frankfurt.de}

\author{Jonathan T.\ Lenaghan}

\address{ Department of Physics \\
	University of Virginia \\
	382 McCormick Rd.	\\
	PO Box 400714\\
	Charlottesville, VA 22904-4714, USA\\
	E-mail: jl8bf@mail.phys.virginia.edu}

\maketitle

\abstracts{ 
The values of renormalized Polyakov loops in the three
lowest representations of
$SU(3)$ were measured numerically on the lattice. We find that in
magnitude, condensates respect the large-$N$ property of
factorization.  In several ways, the deconfining phase transition for
$N=3$ appears to be like that in the $N=\infty$ matrix model of Gross
and Witten.  Surprisingly, we find that the values of the renormalized
triplet loop are described by an $SU(3)$ matrix model, with an effective
action dominated by the triplet loop.  Future numerical simulations
with a larger number of colors
should be able to show whether or not the deconfining phase transition
is close to the Gross-Witten point.}

\section{Introduction} 

't Hooft showed that the order parameter for deconfinement in the
$SU(N)$ Yang-Mills theory is a global
$Z(N)$ spin \cite{zn,polyakov}
given by a thermal Wilson line which wraps around 
the Euclidean time direction.  
The thermal Wilson loop in a representation ${\mathcal R}$ is given by
\be
\boldl_\calr(\vec{x})
	= \calp \; 
	\exp \,\left(  i g \int^{1/T}_0 A_0^a(\vec{x},\tau') \;
	{\bf t}^a_\calr \; d \tau' \right)  \,\, 	 .  
\ee
Under a gauge transformation, 
\bea
\boldl_\calr(\vx) 
\rightarrow \Omega_\calr^\dagger(\vx,1/T)\, \boldl_\calr(\vx)\,
\Omega_\calr(\vx,0) \; .
\eea
There are important aperiodic gauge transformations:
\bea
\Omega_{{\bf N}}(\vx,1/T) = 
e^{i\phi} \; \Omega_{{\bf N}}(\vx,0) 
\;\;\; , \;\;\;
e^{i N \phi} =1 \;.
\eea
The $Z(N)$ charge $e_\calr$ (defined below) 
determines which loops can condense above the deconfinement temperature:
\bea\langle \tr \; \boldl_\calr \rangle = 0 \;\; , \;\;
T < T_d \;\; ,\;\; e_\calr \neq 0 \;
\eea
\bea
\langle \tr \; \boldl_\calr \rangle \neq 0 \;\; , \;\;
T > T_d \;\; , \;\; \forall \; e_\calr \; .
\eea
The trace of the thermal Wilson line is the Polyakov loop \cite{polyakov},
\beq
 \ell_{{\bf \calr}}(\vec{x}) = \frac{1}{d_{\calr}}
{\rm tr}\,\boldl_\calr(\vec{x})~,
\label{def_ellfund}
\eeq
and is gauge invariant.  We divide by the dimension of the
representation, $d_{\calr}$, so that $\ell_{{\bf \calr}}\to1$ as
$T\to\infty$. 

\begin{figure}[ht]
\centerline{\epsfxsize=3.1in\epsfbox{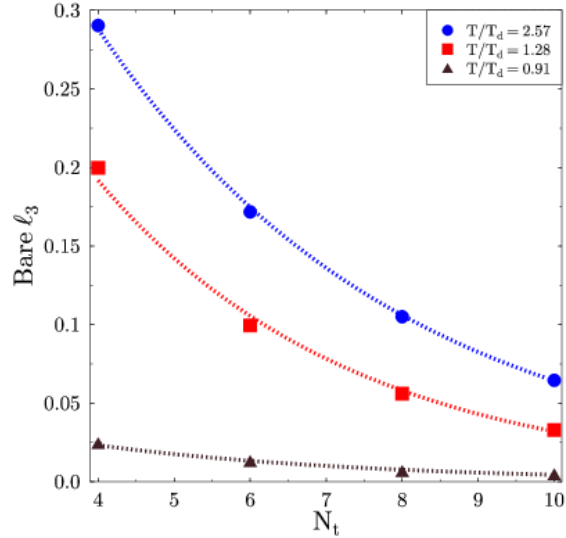}}   
\caption{The bare fundamental loop
 as extracted from lattices with various $N_t$ at three different
temperatures. \label{ContLimit}}
\end{figure}

The low-temperature, confined phase is $Z(N)$ symmetric, so the expectation
value of the fundamental loop vanishes below $T_d$, while above $T_d$
the $Z(N)$ symmetry is broken spontaneously:
\beq
\left\langle \ellfund \right\rangle 
= e^{i \phi} \; \left|\left\langle \ellfund\right\rangle\right| \neq 0
\;\;\; , \;\;\; e^{i \phi N} = 1 \;\;\; , \;\;\; T > T_d \; .
\label{cond_deconf}
\eeq
The physics around $T_d$ is non-perturbative and it is
necessary to employ lattice simulations.  
The expectation value of the Polyakov loop, however, 
is a bare quantity and contains ultraviolet divergences: the
expectation value of the fundamental loop, and of loops in other
representations, vanish in the continuum limit ($N_t\to\infty$), at any fixed
temperature $T=1/(N_t a)$. This is illustrated in Fig.~\ref{ContLimit}.

Renormalization of the expectation
value of the bare Polyakov loop gives the exponential of
a divergent mass times the length of the path
\cite{gervais_neveu,polyakov_ren,ren_other_a,ren_other_b,ren_loop_old,DHLOP}:
\beq
\left|\left \langle \ellfund \right \rangle \right| \sim 
\exp \left(- \frac{m^\sdiv_N}{T} \right) \;\;\; , \;\;\;
m^\sdiv_N \; \sim \; \frac{1}{a} \; .
\label{divergence}
\eeq
The renormalized loop, $\widetilde{\ell}_\calr$,
is formed by dividing the bare loop
by the appropriate renormalization constant, $\calz_\calr$:
\beq
\widetilde{\ell}_\calr = \frac{1}{\calz_\calr} \; \ell_\calr \; \;\; , \;\;\;
\calz_\calr = \exp \left( - \frac{m^\sdiv_\calr}{T} \right) \; .
\label{rendloop}
\eeq

We have developed a method to extract the divergent masses
non-perturbatively~\cite{DHLOP}.  
The idea is to compute with a set of lattices, all
at the same physical temperature, $T$, but with different values of
the lattice spacing, $a$.  The number of time steps, $N_t=1/(aT)$,
varies between these lattices and the divergent mass, $a
m^\sdiv_\calr$, can be extracted by comparing the different values of
the bare Polyakov loops.

An alternate procedure was developed by Kaczmarek, Karsch, Petreczky,
and Zantow \cite{bielefeld_ren} who obtain $\calz_3$ from the two
point function of fundamental loops at short distances.  Our numerical
values for the triplet Polyakov loop agree approximately with their
values. So far, they have not considered loops in other representations.

When the fundamental loop condenses it induces expectation values for 
all loops in higher representations, such as the
octet and sextet loops. 
This can be precisely established in the limit of an infinite number of colors.
Migdal and Makeenko observed that in $SU(N)$ gauge theories,
expectation values factorize at large $N$
\cite{largeN,eguchi_kawai,gocksch_neri,gross_witten,kogut,matrix_deconfining,green_karsch,matrix_review}.
At infinite $N$, factorization fixes
the expectation value of any Polyakov loop to be equal to powers
of those for the fundamental and anti-fundamental,
$\wideellfundconj=(\wideellfund)^*$, loop \cite{eguchi_kawai}:
\begin{equation}
\left\langle \widetilde{\ell}_\calr\right\rangle
= \left\langle \wideellfund \right\rangle^{p_+}
\left\langle \wideellfundconj \right\rangle^{p_-}
+ O\left(\frac{1}{N}\right) \;\;\; . \;\;\;
\label{mean_fieldA}
\end{equation}
Hence
\begin{equation}
\left\langle \widetilde{\ell}_\calr\right\rangle
= e^{i e_\calr\phi} \;
\left|\left\langle \wideellfund \right\rangle\right|^{p}
+ O\left(\frac{1}{N}\right) \; .
\label{mean_fieldB}
\end{equation}
This relation defines the $Z(N)$ charge $e_\calr$ of a loop in a given
representation.
Thus, at infinite $N$, any renormalized loop 
is an order parameter for deconfinement.

To test these large $N$ relationships numerically,
for each loop we define the difference between
the measured loop and its value in the large $N$ limit.  For three
colors, the expectation value of the sextet difference loop is 
\beq
\left \langle \delta \widetilde{\ell}_{ 6}\right \rangle
= \left \langle \widetilde{\ell}_{ 6}\right \rangle -
\left \langle\widetilde{\ell}_{ 3}\right\rangle^{ 2} \; ,
\label{sextet_diff}
\eeq
and that for the octet difference loop is
\beq
\left\langle \delta \widetilde{\ell}_{ 8} \right\rangle
= \left \langle \widetilde{\ell}_{ 8}\right\rangle -
\left|\left\langle \widetilde{\ell}_{ 3}\right\rangle\right|^2 \; .
\label{octet_diff}
\eeq
Note that the difference loops vanish both at $T\to0$ and
at $T\to\infty$. 
If small, they indicate that factorization is 
approximately satisfied.

\section{Matrix Models}

Assuming that Wilson lines form the degrees of freedom
in an effective theory, take as the partition function
\bea
{\mathcal Z} = \int \Pi_{i} \; d\boldlN(i)
\; \exp \left(- {\mathcal S}(\ell_\calr(i)) \right) \;  .
\eea
Here, $i$ labels sites on a spatial lattice.
The character expansion vastly reduces the possible couplings. 
Requiring the action to be $Z(N)$ invariant allows a sum over $Z(N)$
neutral loops:
\bea
N^2 \; \Sigma_i \Sigma_{\calr}^{e_\calr = 0} \; \gamma_\calr \ell_\calr(i) \; \,\, .
\eea
Add nearest neighbor interactions with couplings $\beta_{\calr,\calr'}$:
\bea
{\mathcal S}_\calr = - \frac{N^2}{3}
\Sigma_{i,\hat{n}} \Sigma^{e_\calr + e_{\calr'}=0}_{\calr, \calr'} \;
\beta_{\calr,\calr'} \;
{\rm Re} \; \ell_\calr(i) \ell_{\calr'}(i + \hat{n}) \; .
\eea
Kogut, Snow, and Stone showed~\cite{kogut}
that for $N\ge3$ in mean field approximation this
model has a first order phase 
transition with a latent heat $\sim N^2$ and that 
\bea
|\langle \widetilde{\ell}_{{\bf N}}\rangle| \approx \frac{1}{2}~\mbox{at}~
T = T_d^+ \; .
\eea


\subsection{Three Color Example}

The simplest $N=3$ action includes just
the triplet loop:
\bea {\mathcal S}_3 = - 3 \beta_3 \;
\Sigma_{i,\hat{n}} \; {\rm Re} \; \ellthree(i) \ellthree^*(i + \hat{n}) 
\;\, . 
\eea
Develop a mean field approximation by replacing
all six nearest neighbors (in three space dimensions) by an average value, 
$\ell_0 = \langle \ell_{\bf 3} \rangle$: 
\bea
\calz = \int \; d\boldl_{\bf 3}
\; \exp\left( + 18 \beta_3 \ell_0 {\rm Re} \ell_{\bf 3} \right) 
\equiv \exp(- 9 {\mathcal V}) \;\, .
\eea
The mean field consistency condition is
\bea \ell_0 = - \frac{1}{2 \beta_3}
\; \frac{\partial}{\partial \ell_0} {\mathcal V} \; .
\eea

\begin{figure}[ht]
\vspace*{-1cm}
\centerline{\epsfxsize=3.1in\epsfbox{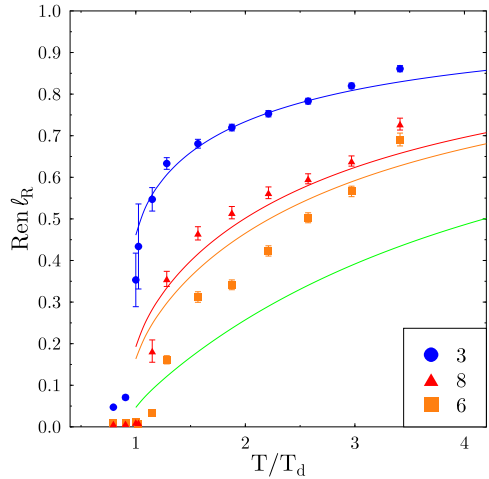}}   
\caption{The condensates computed from the matrix model and compared
  to the lattice data\protect\cite{DHLOP}; for $N=3$. 
The lowest line is the decuplet
  condensate from the $N=3$ matrix model for which no lattice data is
  available.} \label{MMconds}
\end{figure}

We have studied the condensates of the four lowest representations of
$SU(3)$ as a function of the coupling $\beta_3$~\cite{DHLOP}, which
are shown in Fig.~\ref{MMconds}. 
Our first observation is that the ordering (fundamental, octet,
sextet) agrees with that obtained from the lattice.  
A fit of $\langle\ell_3\rangle(\beta_3)$ obtained within the matrix
model to $\langle\ell_3\rangle(T/T_d)$ from the lattice gives
\bea
\beta_3 = (0.46 \pm 0.02) + (0.33 \pm 0.02) \, \frac{T}{T_d} \,\, .
\eea
Remarkably, the matrix model also predicts a sizable expectation
value for the decuplet
 loop above $T_d$; so far, this has not been
studied on the lattice. Also note that expectation values of $Z(3)$
neutral loops (octet
 and decuplet)
 are approximately zero in the
confined phase. This is natural in a matrix model~\cite{DLP}
but does not follow
automatically in an effective theory for a scalar $\ell_3$ 
field~\cite{sv_yaffe,pol_loop_a,PL2,pol_loop_b,sannino}.

\begin{figure}[ht]
\vspace*{-1cm}
\centerline{\epsfxsize=3.1in\epsfbox{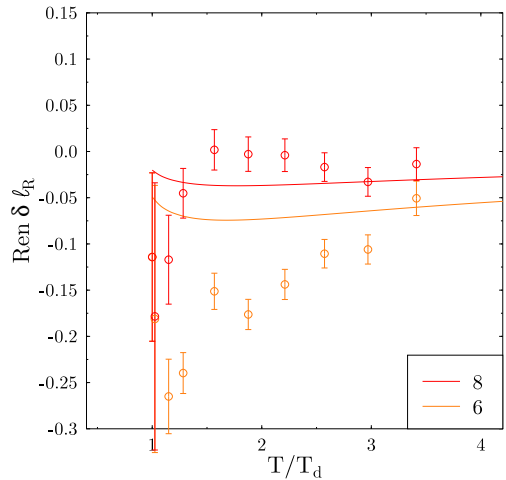}}   
\caption{The difference loops computed from the matrix model and
  compared to the lattice data.} \label{DiffLoops}
\end{figure}

While $\beta_3$ is linear in the
temperature, this doesn't seem to be true for all
couplings. Fig.~\ref{DiffLoops} illustrates this using the difference loops.
Clearly, the ``spikes'' are much smaller and broader in the matrix
model than in 
the lattice data.  From this we conclude that for a more quantitative
interpretation of the lattice results more terms are needed in
the matrix model.

\section{Large--N Matrix Model and the Gross-Witten Point}

At infinite $N$, we do not have to consider loops in higher
representations as independent degrees of freedom.  
In mean--field approximation, we have a single--site 
partition function:
\be
\calz = \int \; d\boldl_{\bf N}
\; \exp\left( N^2 (2 \beta \ell_0) \; {\rm Re} \; \ell_{\bf N} \right)
	\equiv
\exp\left(- N^2 {\mathcal V}_{GW}(\beta \ell_0)\right)\; ,
\ee
where $\beta\equiv \beta_N$.
The mean field condition is
\bea
\ell_0 = - \frac{1}{2 \beta}
\; \frac{\partial}{\partial \ell_0} {\mathcal V}_{GW}(\beta \ell_0) \; ,
\eea
which amounts to minimizing the mean field potential
\bea
{\mathcal V}_{mf}(\beta,\ell) = \beta \ell^2 + {\mathcal V}_{GW}(\beta \ell) \; .
\eea
This has been computed in the large $N$ limit by Gross and 
Witten~\cite{gross_witten}.

The result is nonanalytic and is given by two {\it different}
potentials:
\bea
{\mathcal V}_{mf}^- 
= \beta (1 - \beta) \ell^2 \;\;\; , \;\;\; \ell < \frac{1}{2 \beta} \; \\\nonumber
{\mathcal V}_{mf}^+
= - 2 \beta \ell + \beta \ell^2 + 
\frac{1}{2} \log (2 \beta \ell) + \frac{3}{4} 
\;\;\; , \;\;\; \ell > \frac{1}{2 \beta} \; .
\eea
When $\beta < 1$, the theory confines and $\ell_0 = 0$.
For $\beta \ge 1$ it deconfines and
\bea\nonumber
\ell_0 = 
\frac{1}{2} \left( \; 1 \; + \; 
\sqrt{ \frac{\beta - 1}{\beta} } \; \; \right) \; .
\eea
At $\beta = 1$, the potential is completely {\it flat}, i.e.\ it
vanishes identically for $\ell\in[0,\frac{1}{2}]$.

\subsection{Large--N and Mass of Polyakov Loop}

The connected two point function of $\ell$ is
\bea
\langle \ell^*_{\bf N}(\vec{x}) \ell_{\bf N}(0)\rangle
- |\langle \ell_{\bf N}\rangle|^2 \sim
\frac{\exp(- m |\vec{x}|)}{|\vec{x}|} \;\;\; , \;\;\;
|\vec{x}| \rightarrow \infty \; .
\eea
In the confined phase, $m = \sigma/T$, where $\sigma$ is the string
tension.  In the deconfined phase, one can define $m \propto m_{Debye}$.

Computing $m^2 = \partial^2 {\mathcal V}_{mf}/\partial \ell^2$ gives
\bea
m^2_- \approx 2 (1 - \beta) \;\;\; , \;\;\;
\beta \rightarrow 1^- \; .
\eea
\bea
m^2_+ \approx 4 \sqrt{ \beta - 1} \;\;\; , \;\;\;
\beta \rightarrow 1^+ \; .
\eea
The string tension then vanishes at the transition as
\bea
\sigma(T) \sim (T_d - T)^{1/2}  \; \; \; , \;\;\;
T \rightarrow T_d^- \; ,
\eea
and the Debye mass, as
\bea
m_{Debye}(T) \sim (T - T_d)^{1/4} \; \; \; , \;\;\;
T \rightarrow T_d^+ \; .
\eea
Hence, at the ``Gross-Witten point''~\cite{DHLOP} there is a
``critical'' first order transition where the order parameter jumps
at $T_d$, yet both $m_-$ and $m_+$ vanish.

Numerical results seem to indicate that the three-color Yang-Mills
theory is close to the Gross-Witten point: i) the discontinuity of the
renormalized fundamental loop is approximately 1/2; ii) the ``spikes''
of the difference loops are of order $1/N$ near $T_d$ and vanish at
high $T$; iii) the string tension and the Debye mass drop sharply near
the transition~\cite{string_ten}.

\section{Conclusions}

It would be valuable to know from numerical simulations if the
deconfining transition for more than three colors is close to the
Gross--Witten point as well, or if that is unique to three colors.
This is interesting, novel physics which can be obtained from
lattice measurements at various $N$ of the renormalized Polyakov
loops in hot $SU(N)$ Yang-Mills theory.
Further lattice simulations with higher accuracy will also 
constrain the couplings of the effective matrix model description of
the deconfining phase transition.\\[1cm]
{\bf Acknowledgement:} This work was done in collaboration with
Yoshitaka Hatta, Kostas Orginos and Robert Pisarski. We thank the organizers of SEWM~2004 for
the opportunity to attend this very productive conference.

\end{document}